\documentclass[
aip,
 amsmath,amssymb,
 reprint,%
]{revtex4-1}

\usepackage{graphicx}
\usepackage{dcolumn}
\usepackage{bm}

\usepackage[utf8]{inputenc}
\usepackage[T1]{fontenc}
\usepackage{mathptmx}
\usepackage{etoolbox}

\makeatletter
\def\@email#1#2{%
 \endgroup
 \patchcmd{\titleblock@produce}
  {\frontmatter@RRAPformat}
  {\frontmatter@RRAPformat{\produce@RRAP{*#1\href{mailto:#2}{#2}}}\frontmatter@RRAPformat}
  {}{}
}%
\makeatother
\begin{document}

\preprint{AIP/123-QED}

\title[Modeling scale-up of particle coating by ALD]{Modeling scale-up of particle coating by atomic layer deposition}
\author{Angel Yanguas-Gil}
 \email{ayg@anl.gov}
\author{Jeffrey W. Elam}%
\affiliation{ 
Applied Materials Division, Argonne National Laboratory, Lemont, IL 60439 (USA)
}%

\date{\today}

\begin{abstract}
Atomic layer deposition (ALD) is a promising technique to functionalize particle surfaces for energy applications including energy storage, catalysis, and decarbonization. In this work, we present a set of models of ALD particle coating to explore the transition from lab scale to manufacturing. Our models encompass the main particle coating manufacturing approaches including rotary bed, fluidized bed, and continuously vibrating reactors. These models provide key metrics, such as throughput and precursor utilization, required to evaluate the scalability of ALD manufacturing approaches and their feasibility in the context of energy applications. Our results show that designs that force the precursor to flow through fluidized particles transition faster to a transport-limited regime where throughput is maximized. They also exhibit higher precursor utilization. In the context of continuous processes, our models indicate that it is possible to achieve self-extinguishing processes with almost 100\% precursor utilization. A comparison with past experimental results of ALD in fluidized bed reactors shows excellent qualitative and quantitative agreement.
\end{abstract}

\maketitle

\section{\label{sec:level1}Introduction}

Over the past 20 years, there has been an increasing interest in the use of atomic layer deposition (ALD) for energy and decarbonization applications. Many of these approaches leverage the intrinsic conformality of ALD to functionalize or coat particles\cite{Longrie_fluidizedbedALD_2014,Weimer_ALDparticles_2019,vanOmmen_ALDfluidized_2019}. Two examples are batteries\cite{Koshtyal_ALDbatteriesreview_2022} and catalysis\cite{Cao_ALDcatalysts_2018}: in battery research it has long been known that a few ALD cycles of materials such as Al$_2$O$_3$ on cathode particles improve device stability, preventing issues such as capacity fade\cite{Jung_ALDbatteries_2010}. In catalysis, ALD has been used to engineer catalyst supports\cite{Linblad_ALDcatalysts_1997, Pellin_ALDmembranes_2005}, synthesize catalytic nanoparticles\cite{Lei_ALDcoreshell_2011,Lu_bimetallicparticles_2014}, and prevent catalyst degradation through protective overcoatings\cite{ Feng_ALDPdcatalysis_2011, Lu_ALDcatalysis_2012}.

One of the perceived drawbacks of ALD is that it is a slow process: first, ALD is a pulsed, sequential process, with each ALD cycle involving alternate exposures to two different precursors separated by purge times. Second, surfaces in self-limited processes become less reactive with time as the number of available surface sites decreases. These two factors tend to reduce ALD's throughput compared to non self-limited processes like chemical vapor deposition (CVD).

Intuitively, we understand that, as we scale an ALD process to larger batch sizes and increasingly higher surface areas, the process will eventually transition from a reaction limited to a transport limited regime, thereby eliminating some of the inefficiencies related to the self-limited surface kinetics. Currently, though, we do not have a good understanding of when and how this transition takes place or how it is affected by either the process surface kinetics or the manufacturing approach. This complicates the transfer of ALD processes from the lab to large scale manufacturing.

In this work, we explore the fundamentals of the scale-up of particle coating by ALD. In particular, we focus on three common manufacturing approaches for particle coating: rotating drum\cite{McCormick_rotaryALD_2007,Duan_rotary_2015,Coile_ALDrotarydrum_2020}, fluidized bed\cite{Hakim_particlesALD_2006}, and continuous reactors\cite{vanOmmen_spatialALDparticles_2015,Hartig_CVRmodel_2021}. 
As the interest in ALD manufacturing has
increased, a number of works have tackled particle coating in ALD from a simulation
standpoint\cite{Grillo_multiscalefluidized_2015,Duan_simulations_2017,Jin_fluidizedALDsim_2017,Hartig_particleALD_2023}. These works have usually focused on a single specific
particle coating technique. Our
approach is to generalize traditional chemical engineering models to incorporate the spatial and time dependencies intrinsic to self-limited processes\cite{YanguasGil_diffusionALD_2012}. This allows us to compare the scaling behavior of different ALD particle coating strategies. In a prior work, we showed how in the specific case of a fluidized reactor such simple models provide excellent qualitative and quantitative agreement with experimental data\cite{Lu_fluidizedbedALD_2022}. There is therefore an opportunity to apply the same methodology to other scaling approaches and to expand the range of surface kinetics models considered.

\section{Model}

In this section we introduce the models for the ALD of particles on rotating drum, fluidized bed, and continuous reactors. In Section \ref{sec_kin} we describe the surface kinetic models, while in Section \ref{sec_reactors} we introduce the different reactor models. A list of the symbols used is presented
in Table \ref{tab:symbols}.

\subsection{\label{sec_kin} Surface kinetics}

\subsubsection{Ideal self-limited model}

A key feature of self-limited processes is the presence of a finite number of reactive sites on the surface. Using the open site formalism\cite{Kee_flowbook_2018}, the surface reactivity can be represented by a sticking probability, $\beta$, which depends on the fraction of surface sites available for precursor molecules to react with at a given time. If we define the fractional coverage, $\Theta$, as the fraction of surface sites that have already reacted, the simplest model introduces a first-order dependence on the fraction of available sites, $1-\Theta$: 
\begin{equation}
\label{eq:base}
\beta=\beta_0 (1-\Theta)
\end{equation}

 Where $\beta_0$ represents the reaction probability of the precursor molecule with an available site. This model, which corresponds to a first order irreversible Langmuir kinetics, is one of the most commonly used in the literature to model the surface kinetics of an ALD process\cite{Dendooven_ALDconformality_2009,YanguasGil_simpleALD_2012,Arts_conformalityprobabilities_2019,YanguasGil_reactorsim_2021}.

\begin{table}
\caption{\label{tab:symbols}
List of symbols used in this work}
\begin{ruledtabular}
\begin{tabular}{cl}
Symbol & Interpretation  \\
\hline
$\beta$ & total precursor sticking probability  \\
$\beta_i$ & reaction probability with an available site $i$  \\
$f_i$ & fraction of surface site $i$  \\
$\Theta_i$ & fractional surface coverage of surface site $i$  \\
$s_0$ & average area of a surface site  \\
$J$ & surface flux of precursor molecules \\
$v_{th}$ & precursor mean thermal velocity  \\
$p$ & precursor partial pressure  \\
$n$ & precursor molecular density  \\
$n_0$ & precursor molecular density at inlet  \\
$T$ & process temperature  \\
$V$ & reactor volume \\
$S$ & total particle surface area \\
$t_\mathrm{res}$ & precursor residence time \\
$u$ & flow velocity in plug flow model \\
$S_0$ & reactor cross-section area in plug flow model \\
$L$ & reactor length in plug flow model \\
$v$ & velocity of moving bed of particles \\
$\phi$ & carrier gas volumetric flow \\
$t_0$ & time at which the number of precursor molecules inserted \\
& in the reactor is equal to the number of surface sites \\
$t_s$ & particle residence time in a continuous process \\
$\mathrm{Da}$ & Damköhler number \\
$\gamma $ & excess number, precursor molecule to surface site ratio \\
$\tau$ & normalized dose time for batch models\\
$\tau_s$ & normalized particle residence time for continuous process \\
$\xi$ & normalized reactor length for plug flow models \\
$x$ & normalized precursor concentration \\
\end{tabular}
\end{ruledtabular}
\end{table}

The evolution of the fractional surface coverage is therefore given by:
\begin{equation}
\label{eq:lang}
\frac{d\Theta}{dt} = s_0 \beta_0 (1-\Theta) J
\end{equation}
where $s_0$ represent the average surface area of a reactive site, and $J$ represents the surface flux of precursor molecules (number of precursor molecules reaching the surface per unit area and unit time), which can be calculated from the precursor partial pressure, $p$, and the process temperature, $T$:

\begin{equation}
J=\frac{1}{4} v_{th} n=\frac{1}{4}v_{th}  \frac{p}{k_B T}
\end{equation}
Here $v_{th}$ is the precursor thermal velocity, n is the precursor molecular density (number of molecules per unit volume), and $k_B$ is Boltzmann’s constant.

\subsubsection{Extension to heterogeneous surfaces}

The model above represents the ideal case where the surface can be partitioned into identical reaction sites. However, in many cases of interest surfaces are heterogeneous, and we need to extend Eq. \ref{eq:base} to consider the presence of multiple sites:
\begin{equation}
\beta = \sum_s f_s \beta_s\left(1-\Theta_s\right)
\end{equation}
Here $f_s$ represents the fraction of each site,
$\beta_s$ is the reaction probability of the precursor molecule with that site, and
$\sum_sf_s \le 1$ is the total fraction of reactive surface sites.
In this work we consider the following two cases: in the first scenario only a fraction of the surface is reactive towards the ALD precursor. In this case, if $f$ represents the fraction of the sites that are reactive, we have that:
\begin{equation}
\label{eq:betaf}
    \beta = f \beta_0(1-\Theta)
\end{equation}
resulting in a lower sticking probability, with the evolution of $\Theta$ still given by Eq. \ref{eq:lang}. This scenario is representative of surface functionalization or passivation approaches where nucleation is potentially sparse. Including $f$ in our model allows us to independently tune the size of a surface site, given by its area $s_0$, and the number of active sites on the surface.

The second scenario comprises surfaces with two types of reactive sites, one with high and another with low sticking probability\cite{Bielinski_heatTMA_2022}. This scenario allows us to consider “soft-saturating” ALD processes where, after an initially fast mass uptake, the growth per cycle as a function of dose time asymptotically reaches a limiting value\cite{YanguasGil_reactorsim_2021}. In this scenario, the sticking probability is given by:
\begin{equation}
\label{eq:soft0}
    \beta = f_a \beta_a(1-\Theta_a) +
    f_b \beta_b(1-\Theta_b)
\end{equation}
where $\Theta_a$ and $\Theta_b$ are the fractional coverage of sites a and b, $f_a$ and $f_b$
are the fraction of sites for each reaction pathways, and the total fractional surface coverage is given by:
\begin{equation}
    \Theta = f_a \Theta_a + f_b \Theta_b
\end{equation}
In this case, both $\Theta_a$ and $\Theta_b$
evolve as a function of time following Eq. \ref{eq:lang}.  If $f_a+f_b=1$, we can make $f_b=f$ and 
Eq. \ref{eq:soft0} can be expressed as:
\begin{equation}
\label{eq:soft}
    \beta = (1-f)\beta_a(1-\Theta_a) +
    f \beta_b(1-\Theta_b)
\end{equation}

\subsection{\label{sec_reactors}Reactor models}

\begin{figure}
\includegraphics[width=8cm]{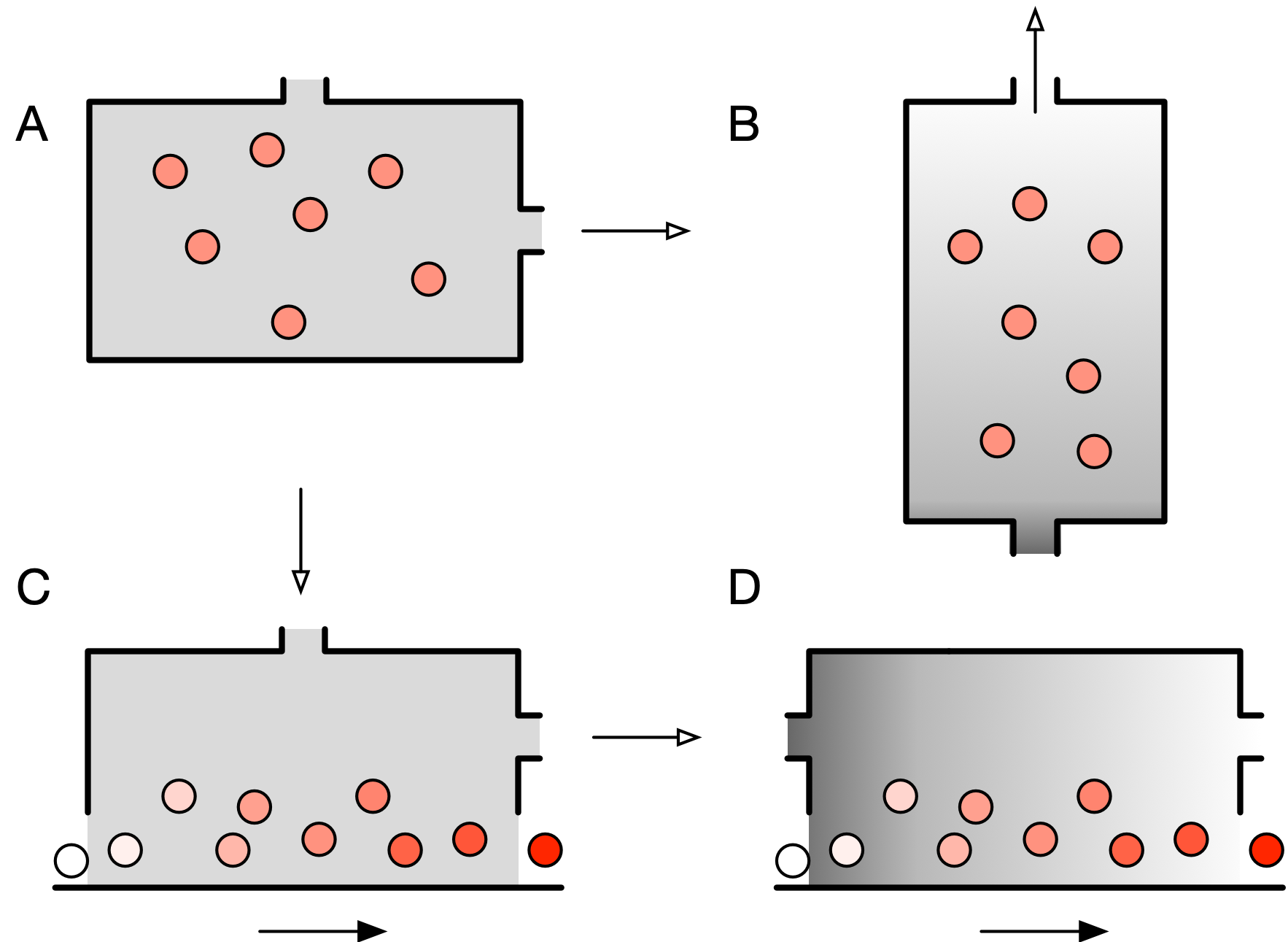}
\caption{\label{fig1} Four type of reactor models considered in this work: A) well-mixed batch process; B) plug flow batch process; C) well-mixed continuous process; D) plug flow continuous process.  In these diagrams, the white arrows show the direction of precursor flow, and the black arrows designate the direction of particle transport.  The particle shading represents the degree of saturation with dark red indicating complete coverage.  The background shading indicates precursor concentration with darker shading representing a higher concentration.}
\end{figure}

There are two key characteristics 
that we can use to classify different particle coating strategies. We will use them to build four different
models for ALD particle coating that cover all possible combinations (Fig. \ref{fig1}).

A first distinction is whether
particles are coated in batches or are continuously fed into the reactor. Examples of batch processes
include rotating drum and fluidized bed reactors. For the batch models, we assume homogeneous
particle mixing. This corresponds to experimental configurations in which the characteristic
dose time is much larger than the characteristic particle mixing time. In this regime, all particles
are homogeneously coated as they sample conditions all over the reactor.

In continuous processes, particles are fed into the reactor continuously, being exposed to the ALD
precursor as they travel inside the reactor [Figs. \ref{fig1}(C) and (D)].
Examples of continuous processes include continuous vibrating reactors (CVR) and continuous
fluidized reactors. For continuous processes, we assume that particles are fluidized or agitated
in a way that achieves perfect mixing in directions perpendicular to particle transport\cite{Hartig_CVRmodel_2021,Hartig_particleALD_2023}. As a consequence, the surface coverage
of particles increases as particles move further along the reactor. The velocity of the moving particles
$v$ and the length of the reactor $L$ are key variables in continuous processes, as they 
determine the particle residence time inside the reactor.
Due to the perfect mixing, particles
exiting the reactor all have identical surface coverage. In both the batch and continuous cases, a narrow distribution of film thickness is a necessary condition for these
approximations to hold.

A second distinction involves precursor transport inside the reactor: when the geometry and
configuration of the reactor favors precursor mixing, the precursor mixing time is much faster than
the characteristic dose time. Under these conditions,
we can model precursor transport using the well-mixed approximation [Fig. \ref{fig1}(A)],
and the reactor volume $V$ and characteristic residence time $t_\mathrm{res}$ are the key
variables defining precursor transport. Rotating drum reactors and configurations that force precursor
mixing are 
examples of processes where the well-mixed approximation can hold.

Alternatively, the reactor can be designed to force precursor to flow through it. This
is the case of of some fluidized bed reactors with tall and narrow fluidized beds [Fig. \ref{fig1}(B)].
Prior works involving
ALD on both cross-flow reactors and particle coating in fluidized bed reactors have shown that plug
flow models of precursor transport reproduced well the experimental results\cite{Lu_fluidizedbedALD_2022,
Yanguas-Gil_analytic_2014}. In reactors where precursor transport can be approximated by a plug flow model,
the flow velocity $u$, the cross-sectional area of the reactor $S_0$, and the reactor length $L$ are
key variables.

The well-mixed and plug flow approximations can be viewed as two limiting cases for precursor transport that can be applied to both batch and continuous ALD processes. Consequently, in this work we explore all four possible combinations,
as shown in Fig. \ref{fig1} and Table \ref{tab:models}.
We should emphasize that the mapping between these four models and strategies for particle coatings is not one-to-one: for instance, depending on the reactor geometry, continuous vibrating reactors can
be modeled using the well-mixed and the plug flow approximations
[Figs. \ref{fig1}(C) and \ref{fig1}(D), respectively].
Likewise, fluidized bed reactors can be designed to operate in batch or continuous mode, and rotating drum reactors can have geometries that favor mixing or that transport the precursor along the axis of revolution. Some of the possible combinations are enumerated in Table \ref{tab:models}.

\begin{table*}
\caption{\label{tab:models}
Summary of models and conditions explored in this work}
\begin{ruledtabular}
\begin{tabular}{ccccc}
Model & Process type & Particle mixing
 & Precursor transport & Examples \\
\hline
A & Batch & Homogeneous	& Well-mixed & Rotating drum \\
B & Batch & Homogeneous & Plug flow & Fluidized bed, rotating drum \\
C & Continuous & In plane mixing & Well-mixed & CVR, spatial ALD \\
D & Continuous & In plane mixing & Plug flow &	CVR, fluidized bed, spatial ALD\\
\end{tabular}
\end{ruledtabular}
\end{table*}

In all these cases, the particle loading is determined by the total surface area $S$ of the particles inside the reactor. When particle loadings are high, particles can occupy a non-negligible fraction $\varepsilon$ of the 
particle volume $V$ and the reactor cross-sectional area $S_0$. These factors can be incorporated
into the model constants.  Finally, the number
of precursor moles inserted in the reactor is proportional to the precursor molecular density $n_0$ or,
equivalently to the precursor partial pressure $p_0$, at the inlet. These will be used as initial
conditions in all the models considered in this work.

\subsubsection{\label{sec:modelA}Batch process with well-mixed reactor approximation}

This model assumes ideal mixing so that all particles are homogeneously coated [Figure \ref{fig1}(A)]. 
Under the assumption that all particles are homogeneously mixed, we only
need to consider a single value for the fractional surface coverage.
Furthermore, the precursor is also efficiently mixed in the reactor, and the well-mixed approximation applies.  A prototypic example would be a rotating drum reactor under near-static precursor flow conditions. As mentioned in Section \ref{sec_reactors}, the precursor transport is defined by a residence time $t_\mathrm{res}$. 

For a single reactive pathway, the model equations are:
\begin{eqnarray}
\frac{V}{t_\mathrm{res}} (n_0-n)  & = & S f \beta_0 (1-\Theta) \frac{1}{4} v_{th} n \\
\frac{d \Theta}{dt} & = & s_0 \beta_0 (1-\Theta) \frac{1}{4} v_{th} n
\end{eqnarray}
Whereas for a soft-saturating system (Eq. \ref{eq:soft}) the model equations are:
\begin{equation}
\begin{array}{ccl}
\dfrac{V}{t_\mathrm{res}} (n_0-n) & = & S \left[(1-f) \beta_a (1-\Theta_a) \right.\\
& & \left.+ f\beta_b(1-\Theta_b) \right]
\dfrac{1}{4} v_{th} n \\
\end{array}
\end{equation}
\begin{eqnarray}
\frac{d \Theta_a}{dt} & = & s_0 \beta_a (1-\Theta_a) \frac{1}{4} v_{th} n \\
\frac{d \Theta_b}{dt} & = & s_0 \beta_b (1-\Theta_ b) \frac{1}{4} v_{th} n
\end{eqnarray}

\subsubsection{\label{sec:modelB}Batch process with plug flow precursor transport}

We next consider a plug flow model for the precursor transport across a homogeneously mixed particle column of cross section area $S_0$ and length $L$ [Figure \ref{fig1}(B)].  Examples for this type of reactor include a fluidized bed or
a rotating drum reactor under cross-flow conditions. In this case, the precursor transport is modeled
using a plug flow model:
\begin{equation}
    S_0 u \frac{d n}{dz}  =  -\frac{S}{L} f \beta(t) \frac{1}{4} v_{th} n
\end{equation}
where $u$ is the flow velocity and $z$ is the distance from the precursor inlet.

Under the assumption that the characteristic time for particle mixing is much smaller than the
dose times required to achieve saturation, particles are equally likely to be at any position within
the reactor during a precursor dose. Consequently, they will be exposed to the average precursor
concentration in the reactor at any given time, and the surface kinetics will depend
on the average precursor density $\bar{n}$:
\begin{equation}
\bar{n} = \frac{1}{L}\int_0^Ln(z) dz 
\end{equation}
This leads to
the following equations for the single reaction
pathway case:
\begin{eqnarray}
S_0 u \frac{d n}{dz} & = & -\frac{S}{L} f \beta_0 (1-\Theta) \frac{1}{4} v_{th} n \\
\frac{d \Theta}{dt} & = & s_0 \beta_0 (1-\Theta) \frac{1}{4} v_{th} \bar{n} \\
\bar{n} & =&  \frac{1}{L} \int_0^L n(z) dz
\end{eqnarray}

The following equations are obtained for
the soft-saturating ALD case:
\begin{equation}
\begin{array}{ccl}
S_0 u \dfrac{d n}{dz} & = & -\dfrac{S}{L} \left[(1-f) \beta_a (1-\Theta_a)\right. \\
& & \left. + f\beta_b(1-\Theta_b) \right]\dfrac{1}{4} v_{th} n 
\end{array}
\end{equation}
\begin{eqnarray}
\frac{d \Theta_a}{dt} & = & s_0 \beta_a(1-\Theta_a) \frac{1}{4} v_{th} \bar{n} \\
\frac{d \Theta_b}{dt} & = & s_0 \beta_b (1-\Theta_b) \frac{1}{4} v_{th} \bar{n}
\end{eqnarray}

\subsubsection{\label{sec:modelC}Continuous process and well-mixed precursor approximation}

We next consider the case of a continuous process. As described in Section \ref{sec_reactors},
we assume that particles move with a velocity $v$ inside a reactor of length $L$ [Figure \ref{fig1}(C)]. 
The particles are only partially mixed, with perfect mixing assumed only in the plane perpendicular to the direction of movement. 
Consequently, the surface coverage will depend on the particle position along the upstream-downstream axis
$z$, $\Theta = \Theta(z)$, and on the total exposure received until reaching $z$. 
Furthermore, in this model we assume that the precursor is well-mixed inside the reactor.  An example for this type of system would be a vibro-fluidized bed reactor under near static flow conditions.

Under the well-mixed approximation, the precursor density is given by a balance between the moles
of precursor coming into the reactor at any given time and the precursor losses due to transport out of the
reactor and the heterogeneous processes. The total precursor loss to surface reactions can
be calculated by integrating the sticking probability $\beta$ along the upstream-downstream axis.

For a single heterogeneous process, this leads to the following equations:
\begin{eqnarray}
\frac{V}{t_\mathrm{res}} (n_0-n)  & = & S f \beta_0 (1-\bar{\Theta}) \frac{1}{4} v_{th} n \\
v \frac{d \Theta}{dz} & = & s_0 \beta_0 (1-\Theta) \frac{1}{4} v_{th} n
\end{eqnarray}
Where $\bar{\Theta}$ is the average reactivity along the upstream downstream axis:
\begin{equation}
    \bar{\Theta}  =  \frac{1}{L}\int_0^L \Theta(z) dz
\end{equation}
This model is very similar to a model we previously developed for spatial ALD\cite{Yanguas-Gil_analytic_2014}, except that in this case the surface area is dominated by the total particle surface
area $S$.

This model is trivially extended to the soft-saturating case:
\begin{equation}
\begin{array}{ccl}
\dfrac{V}{t_\mathrm{res}} (n_0-n) & = & S \left[(1-f) \beta_a (1-\bar{\Theta}_a) \right.\\
& & \left.+ f\beta_b(1-\bar{\Theta}_b) \right]
\dfrac{1}{4} v_{th} n \\
\end{array}
\end{equation}
and
\begin{eqnarray}
v\frac{d \Theta_a}{dz} & = & s_0 \beta_a (1-\Theta_a) \frac{1}{4} v_{th} n \\
v\frac{d \Theta_b}{dz} & = & s_0 \beta_b (1-\Theta_ b) \frac{1}{4} v_{th} n
\end{eqnarray}
with:
\begin{eqnarray}
    \bar{\Theta}_a & =&   \frac{1}{L}\int_0^L \Theta_a(z) dz \\
\bar{\Theta}_b & =&   \frac{1}{L}\int_0^L \Theta_b(z) dz
\end{eqnarray}

\subsubsection{\label{sec:modelD}Continuous process with plug flow precursor transport}

The final case considered in this work comprises a continuous process in which the precursor is injected in an upstream position and flows downstream over a fluidized bed of particles moving in the same direction [Figure \ref{fig1}(D)]. In this case, both the fractional surface coverage and the precursor concentration depend on the axial position $z$. For the case of a single heterogeneous process:
\begin{eqnarray}
S_0 u \dfrac{du}{dz} & = & -\dfrac{S}{L} f \beta_0 (1-\Theta) \frac{1}{4} v_{th} n \\
v \frac{d \Theta}{dz} & = & s_0 \beta_0 (1-\Theta) \frac{1}{4} v_{th} n
\end{eqnarray}
which again can be trivially extended to the soft-saturating case:
\begin{equation}
\begin{array}{ccl}
S_0 u \dfrac{d n}{dz} & = & -\dfrac{S}{L} \left[(1-f) \beta_a (1-\Theta_a)\right. \\
& & \left. + f\beta_b(1-\Theta_b) \right]\dfrac{1}{4} v_{th} n 
\end{array}
\end{equation}
\begin{eqnarray}
v \frac{d \Theta_a}{dz} & = & s_0 \beta_a (1-\Theta_a) \frac{1}{4} v_{th} n \\
v \frac{d \Theta_b}{dz} & = & s_0 \beta_b (1-\Theta_b) \frac{1}{4} v_{th} n
\end{eqnarray}
Here $v$ is the velocity of the particles, as defined for the continuous well-mixed model, 
$S_0$ is the cross-sectional area of the reactor and $u$ is the flow velocity.

\subsection{Model normalization}

In order to establish a comparison between the models introduced in Section \ref{sec_reactors}, we will use two common non-dimensional numbers: the first is the Damköhler number, which can be defined as the ratio of the characteristic reaction rate and the transport rate:
\begin{equation}
    \mathrm{Da} = \frac{\text{reaction rate}}{\text{transport rate}}
\end{equation}
For the case of a single heterogeneous process, we define the Damköhler number as:
\begin{equation}
\label{eq:Da}
\mathrm{Da} = \frac{S}{V} \beta_0 \frac{1}{4}v_{th} t_\mathrm{res}
\end{equation}
In the case of plug-flow models, the residence time is defined as $t_\mathrm{res}=L/u$. We can introduce the
case of sparse nucleation ($f<1$) by defining an effective
surface area $S_\mathrm{eff} = fS$.
For the soft-saturating case, we can define a Damköhler number for each surface reaction pathway,  $\mathrm{Da}_i$.

The second non-dimensional number is a characteristic time that depends on the precursor residence time and the number of precursor molecules in the reactor per surface site, which in the context of the ALD has been referred to as the excess number, $\gamma$: 
\begin{equation}
    t_0= t_\mathrm{res}\frac{S}{s_0 n_0 V}=\frac{t_\mathrm{res}}{\gamma}
\end{equation}
For a batch process, we can define a normalized dose time in terms of $t_0$:
\begin{equation}
\tau = \frac{t}{t_0}
\end{equation}
The meaning of $\tau$ is that $\tau=1$ corresponds to the time at which the total number of precursor molecules inserted in the reactor is equal to the total number of reactive sites.

For a continuous process, we define a normalized particle residence time:
\begin{equation}
\label{eq:normres}
\tau_s = \frac{t_s}{t_0}
\end{equation}
where $t_s$ is the average particle residence time inside the reactor $t_s=L/v$. As before, $\tau_s=1$ corresponds to the point at which precursor molecules and surface sites are introduced at equal rates.

Finally, we define a normalized reactor length, $\xi = z/L$, and a normalized precursor concentration, $x=n/n_0$. 

The resulting non-dimensional equations for the four models in both the single heterogeneous process case and the soft-saturation case are provided in the Appendix.

\section{Results}

\subsection{\label{sec:solutions}Model solutions}

The four models introduced above can be solved exactly for the case of a single reaction pathway described by Eq. \ref{eq:betaf}.

For a batch process, the fractional coverage $\Theta$ of the well-mixed and plug flow models are given in terms of the normalized dose time
$\tau$ and the Damköhler number, $\mathrm{Da}$, by:
\begin{equation}
\label{eq:modela}
\tau = \Theta - \frac{1}{\mathrm{Da}}\log\left(1-\Theta\right)
\end{equation}
and
\begin{equation}
\label{eq:modelb}
\Theta = 1-\frac{1}{\mathrm{Da}}\log\left(
1+(e^\mathrm{Da}-1)e^{-\mathrm{Da}\tau}\right)
\end{equation}
respectively.

Interestingly, the expression for the final surface coverage for the continuous well-mixed process is identical to the batch well-mixed process case (Eq. \ref{eq:modela}), except that it depends on the normalized residence time $\tau_s$ instead:
\begin{equation}
\label{eq:modelc}
\tau_s = \Theta - \frac{1}{\mathrm{Da}}\log\left(1-\Theta\right)
\end{equation}

Finally, the surface coverage for the continuous plug flow model is given by:
\begin{equation}
    \label{eq:modeld}
    \Theta = 1-\frac{1-\tau_s}{1-\tau_se^{-(1-\tau_s)\mathrm{Da}}}
\end{equation}

\subsection{\label{sec_batch}Scale up of batch processes}

The value of $\mathrm{Da}$ increases with increasing amount of surface area inside the reactor. Consequently, we can use this parameter to track the behavior of a process upon scale up.  An ideal process would achieve saturation in the minimum possible time and utilize 100\% of the supplied precursor.  As will be shown below, all of the reactor models approach ideal behavior in the limit of high $\mathrm{Da}$, but the degree of ideality varies among the models. This highlights the differences in their scalability.

\begin{figure}
\includegraphics[width=8cm]{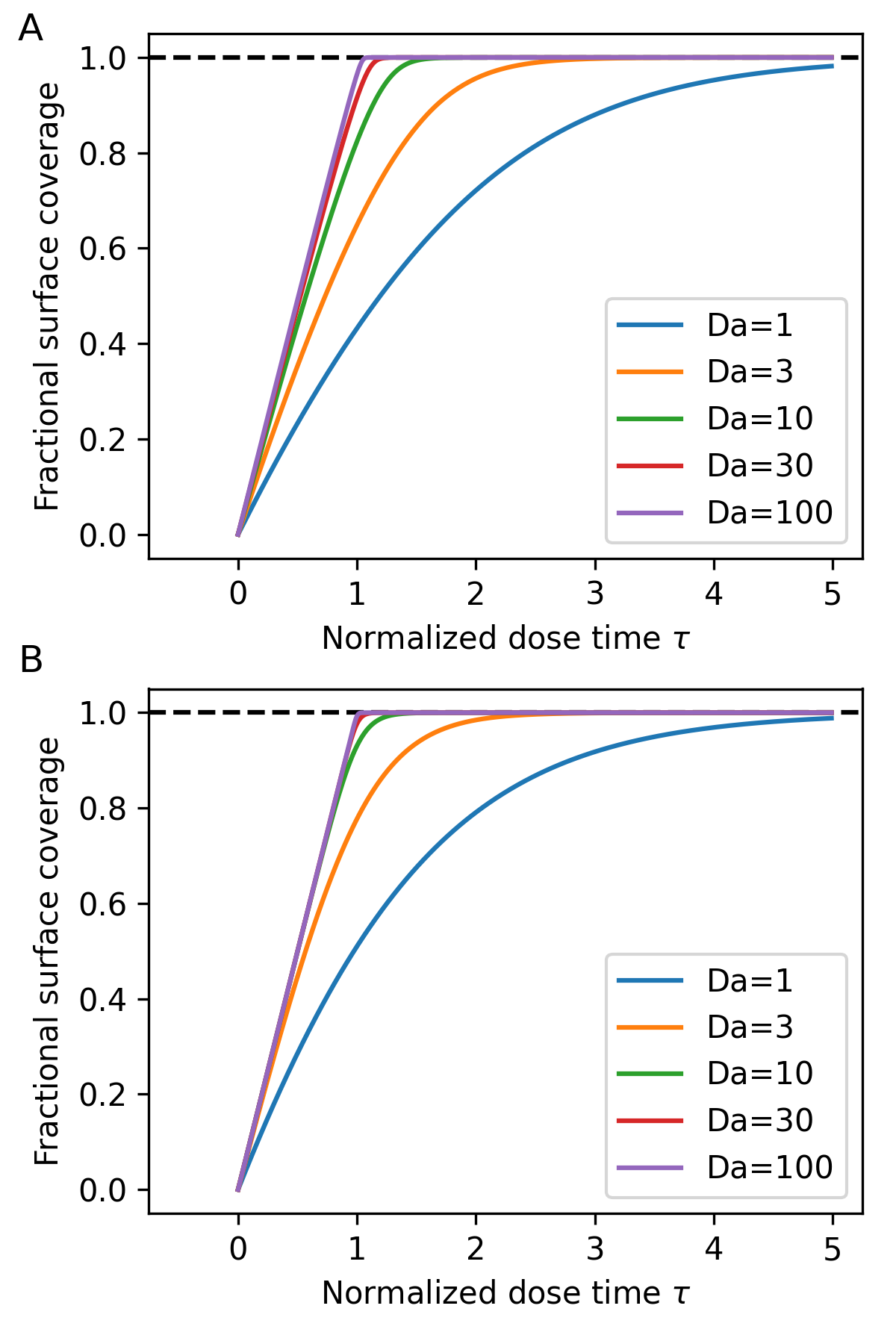}
\caption{\label{fig2} Saturation curves showing the fractional surface coverage of particles as a function of the normalized dose time for the (A) well-mixed batch and (B) plug flow batch models. As the Damköhler number increases, both processes transition from a reaction limited to a transport limited regime.}
\end{figure}

Figure \ref{fig2} shows saturation plots depicting the evolution of the fractional surface coverage as a function of dose time for batch processes with the well-mixed [Fig \ref{fig2}(A)] and plug flow [Fig \ref{fig2}(B)] precursor transport approximations. Results are shown for increasing values of the Damköhler number.
The plots are represented using the normalized dose time $\tau$. A value of $\tau=1$ represents the time at which the total number of moles of precursor inserted in the reactor equals the total number of surface reactive sites, so a fractional surface coverage approaching one for $\tau=1$ indicates that the process is reaching the transport-limited regime and approaching ideal behavior. For both the well-mixed and plug flow models, the slope of fractional coverage vs dose time increases with increasing values of $\mathrm{Da}$. The time to saturation also shortens, reaching values close to $\tau=1$, thus confirming that ALD processes do become more efficient as they are scaled. However, that transition takes place at lower Damköhler numbers for the plug flow model, with the transition taking place for $\mathrm{Da}>10$, whereas for the well-mixed reactor the transition requires
$\mathrm{Da}>30$.

\begin{figure}
\includegraphics[width=8cm]{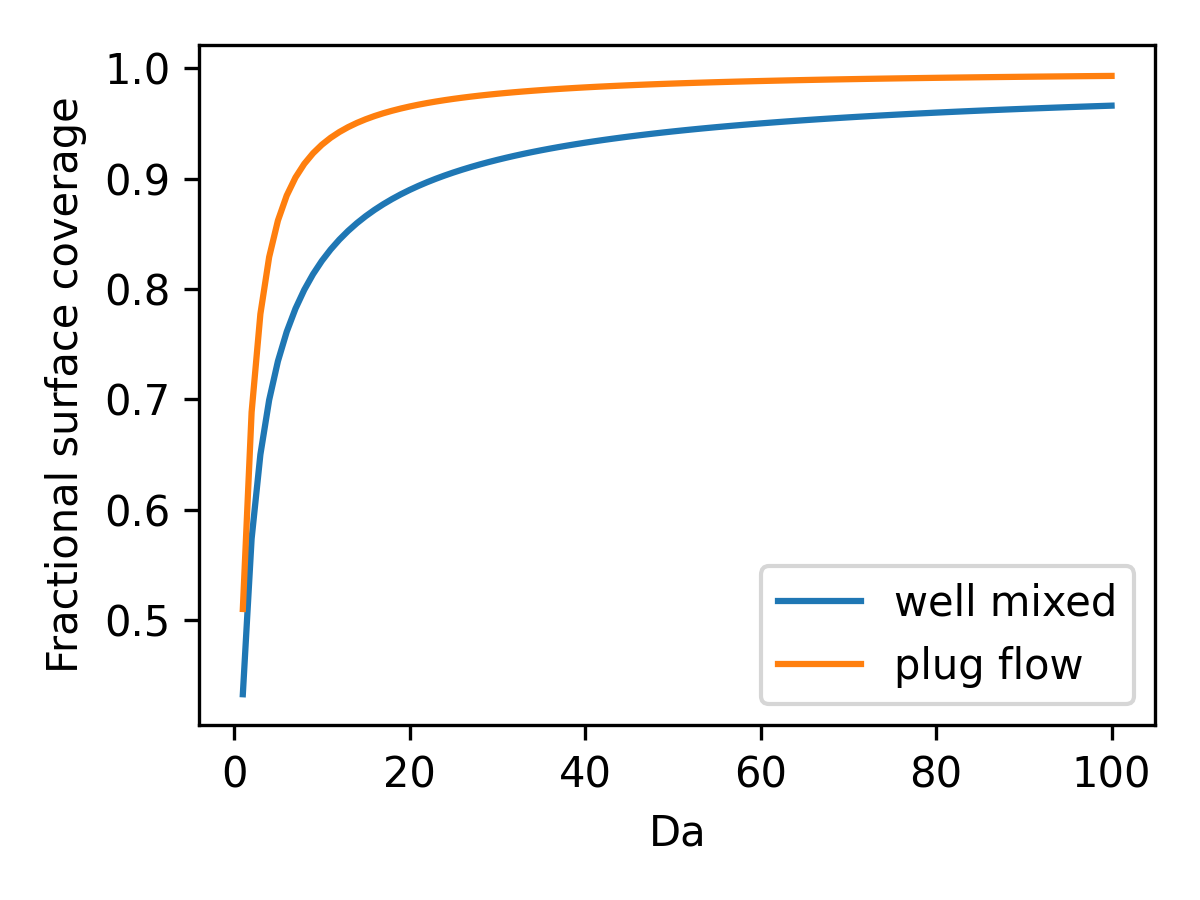}
\caption{\label{fig3} Fractional surface coverage of particles coated by ALD in a batch process for a normalized dose time $\tau=1$ as a function of the Damköhler number. Reactors whose precursor transport can be approximated by a plug flow model transition faster from a reaction limited to a transport limited regime compared to the well-mixed model, characterized by a saturation time equal to $\tau=1$. }
\end{figure}

One way of visualizing the transition from a reaction to a transport-limited regime is by tracking the fractional surface coverage for $\tau=1$ as a function of the Damköhler number. This is shown in Figure \ref{fig3} for both the well-mixed and plug flow models. The results show that the plug flow model transitions faster to a reaction limited regime, achieving larger fractional surface coverages than the well-mixed model for the same process and surface area (equal $\mathrm{Da}$).

\begin{figure}
\includegraphics[width=8cm]{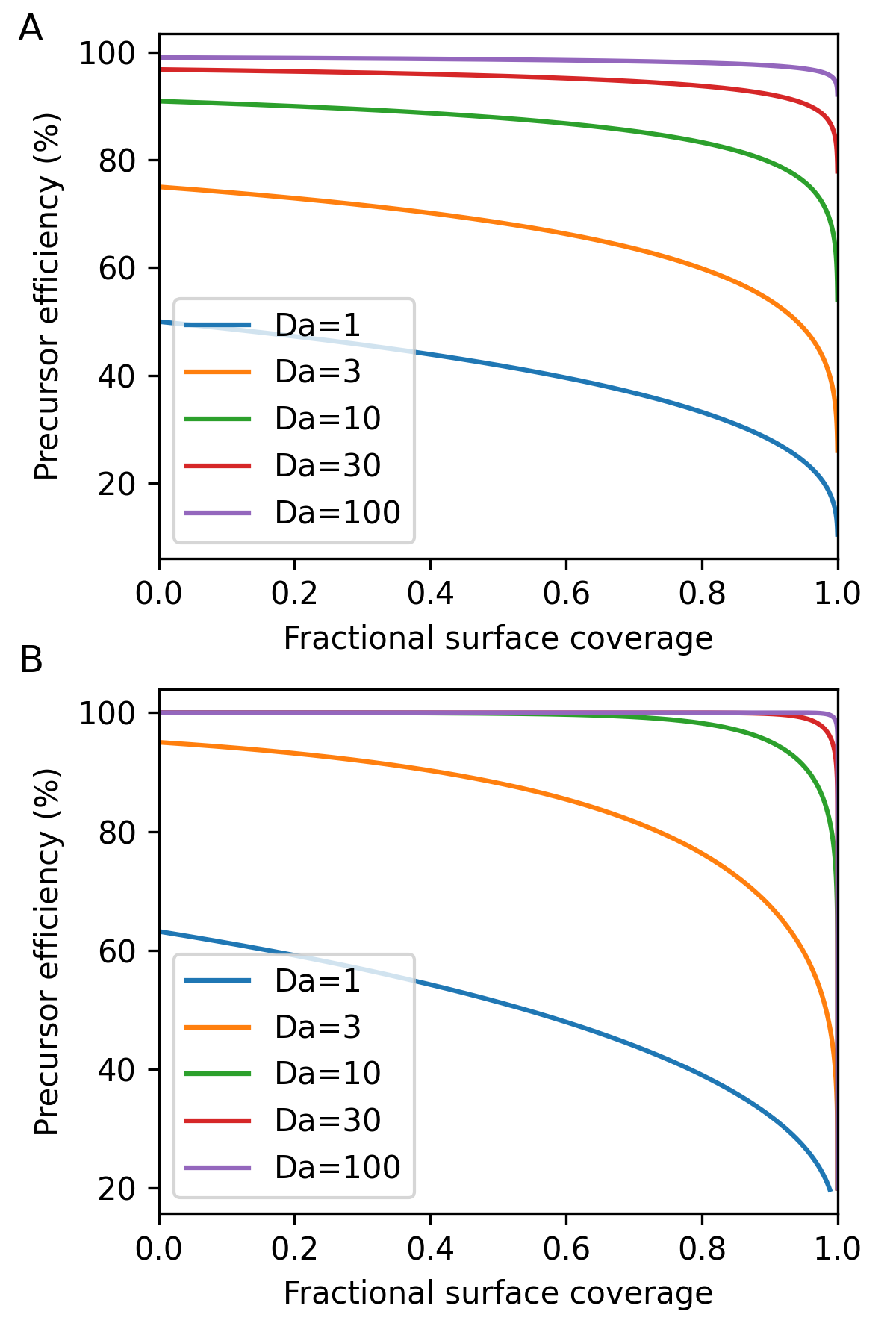}
\caption{\label{fig4} Fraction of precursor used for coating particles as a function of the final target fractional surface coverage for (A) a well-mixed batch and (B) a plug flow batch process. Both processes become more efficient with increasing Damköhler number, with the plug flow batch process outperforming the well-mixed batch process. }
\end{figure}

A crucial metric for manufacturing is precursor efficiency, which we define as the percentage of precursor molecules that are used in the deposition process. In Figure \ref{fig4} we show the precursor efficiency for both the well-mixed [Fig. \ref{fig4}(A)] and plug flow [Fig. \ref{fig4}(B)] models as a function of the final fractional surface coverage for increasing values of the Damköhler number. At low Damköhler numbers, the precursor utilization is low due to the intrinsic slow down of the surface kinetics in self-limited processes. However, as the Damköhler number increases, the process becomes increasingly more efficient until reaching almost 100\% precursor utilization. As shown above for the fractional surface coverage, the plug flow model yields a higher precursor efficiency than the well-mixed model for identical fractional surface coverage and Damköhler number.

\begin{figure}
\includegraphics[width=8cm]{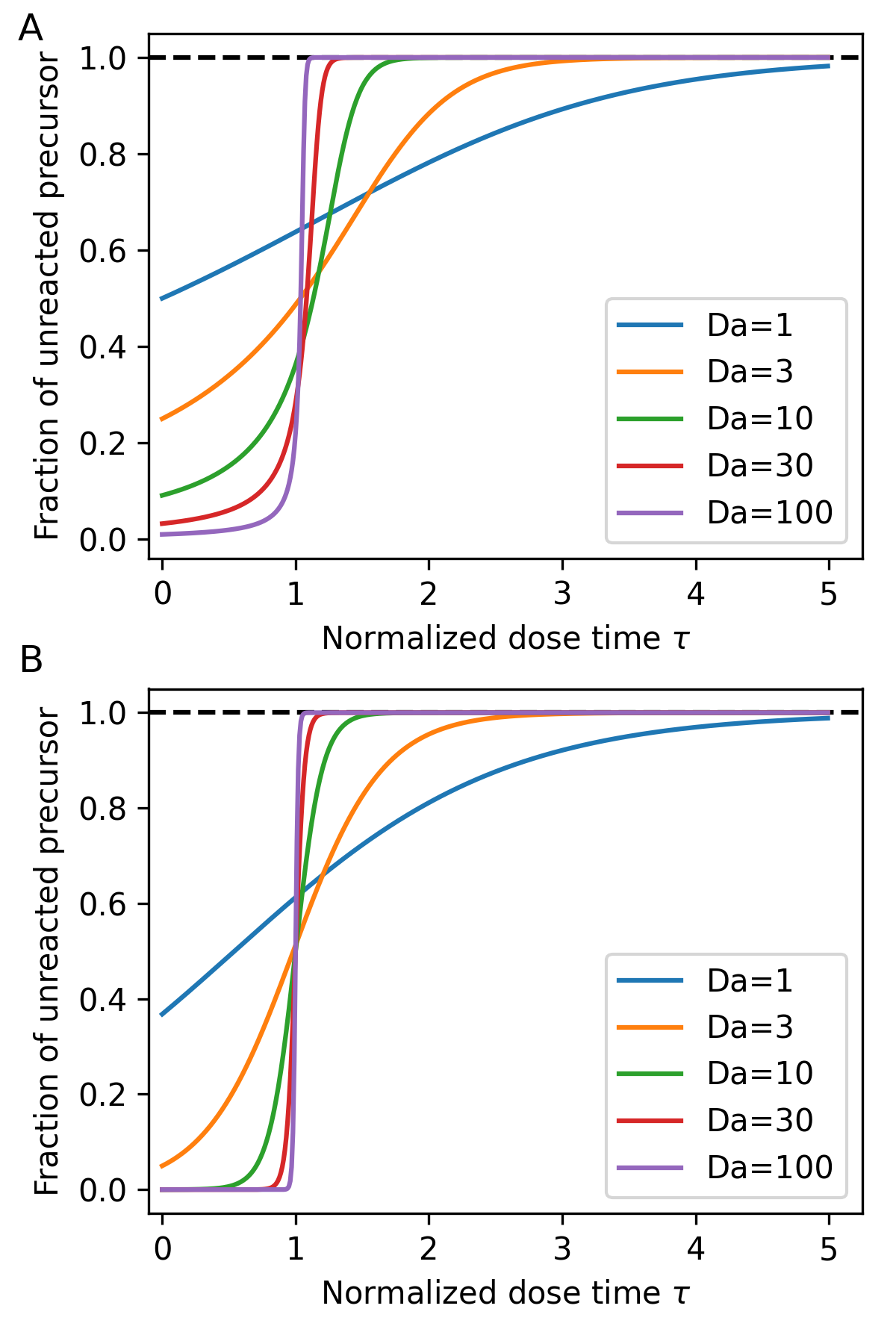}
\caption{\label{fig5} Fraction of unreacted precursor for the (A) well-mixed batch and (B) plug flow batch models as a function of normalized dose time and increasing Damköhler numbers. Under the precursor transport limited regime, a clear signature from the precursor is observed as the ALD process reaches saturation. }
\end{figure}

Likewise, in Figure \ref{fig5} we show the fraction of unreacted precursor that leaves the reactor as a function of normalized dose time for both batch processes. This depiction of precursor efficiency is useful because it models a downstream measurement of the precursor concentration using a mass spectrometer. The high precursor efficiencies shown in Figure \ref{fig4} consistently lead to almost no precursor leaving the reactor prior to reaching saturation. However, as the system transitions to a transport limited regime, the “punch-through” signal of unreacted precursor becomes increasingly more sharp at the point where saturation is reached. 

The model results agree well with experimental observations on fluidized bed reactors: first, in-situ
mass spectrometry provides a characteristic punch-through signal similar to that shown in Fig. \ref{fig5}\cite{Weimer_ALDparticles_2019,Lu_fluidizedbedALD_2022}. In a prior work
focused on Al$_2$O$_3$ ALD from TMA/H$_2$O ALD, we also observed
a similarity between the precursor pressure traces observed at the inlet and the mass
spectrometry signals from reaction byproducts in the downstream position\cite{Lu_fluidizedbedALD_2022}. This is consistent with the 100\% precursor efficiency
expected at high Damköhler numbers (Fig. \ref{fig4}). Finally, in that same work 
we observed that the saturation times, defined as the time at which punch-through was experimentally observed, also agreed well with the predictions from Eq. \ref{eq:modelb}. This indicates that, despite their simplicity, the models can capture the main features of particle coating for ALD, at least in the case of batch fluidized bed reactors.

\subsection{Scale up of continuous processes}

Continuous processes differ from batch processes in two ways: first, in a continuous process the precursor is constantly dosed into the reactor volume so it is the residence time of the particles inside the reactor, $t_s$, what determines the total exposure. Second, the fractional surface coverage of the particles increases as particles move inside the reactor. The relevant observable of a continuous process is therefore the fractional surface coverage as particles exit the reactor. The exit point is defined by the reactor length $L$.

As mentioned in Section \ref{sec:solutions}, in the case of an ideal self-limited process, the solution for the continuous well-mixed model (Eq. \ref{eq:modelc}) is identical to that of the well-mixed batch model, except that the dose time, $\tau$, is replaced with the normalized particle residence time, $\tau_s$
(Eq. \ref{eq:normres}). The results shown in Figures \ref{fig2}(A) and \ref{fig4}(A) are therefore applicable to the well-mixed continuous model.

\begin{figure}
\includegraphics[width=8cm]{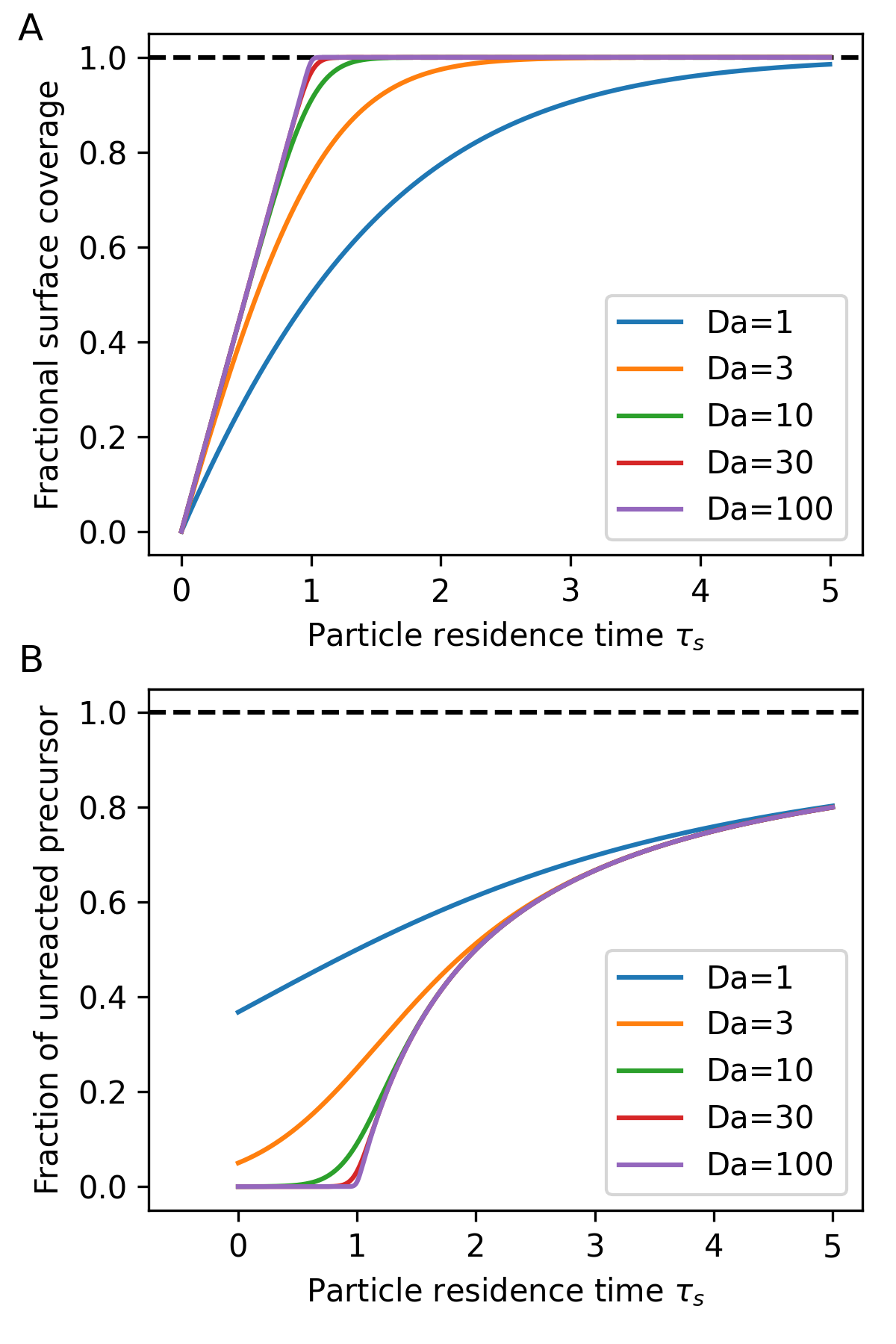}
\caption{\label{fig6} (A) Saturation curves showing the fractional surface coverage of particles as a function of the normalized particle residence time and increasing Damköhler numbers for a continuous plug flow model for particle coatings by ALD. (B) Fraction of unreacted precursor for the same conditions. }
\end{figure}

The key features of the plug flow continuous model are shown in Figure \ref{fig6}. In Figure \ref{fig6}(A) we show saturation curves for increasing Damköhler numbers as a function of the normalized particle residence time, $\tau_s$. The normalized particle residence time is inversely proportional to the velocity at which the particles are being moved into the reactor, and proportional to the reactor size (Eq. \ref{eq:normres}).
  When particles are inserted in the reactor at a very high rate ($\tau_s \rightarrow 0$), particles leave the reactor largely uncoated. 
$\tau_s=1$ corresponds to the case where the number of surface sites and the number of precursor molecules inserted in the reactor per unit time are the same. Consequently, as in the case of batch processes, a linear increase of surface coverage with residence time that leads to saturation at $\tau_s=1$ indicates that the system is in the transport-limited regime.

In Figure \ref{fig6}(B) we show the fraction of unreacted precursor for the same parameters used in Figure \ref{fig6}(A). At very low residence times, the fraction of precursor leaving the reactor is very small, indicating
almost 100\% precursor utilization. For  $\tau_s > 1$, the rate at which precursor is fed into the reactor
exceeds that of the surface sites, and the fraction of unreacted precursor starts to increase. Both Figs \ref{fig6}(A) and \ref{fig6}(B) show that, at high Damköhler numbers, there is a clear transition at $\tau_s=1$ separating the undersaturated regime, where growth is still not fully saturated and precursor is fully
utilized, and the saturated regime, where the surface fractional coverage is very high and precursor utilization starts to decrease. Consequently, like in the batch plug flow case, techniques such as downstream mass spectrometry should provide a clear signature indicating the optimal particle insertion rate at which saturation is achieved. However, this transition is less abrupt than in batch processes (Fig. \ref{fig5}).

\begin{figure}
\includegraphics[width=8cm]{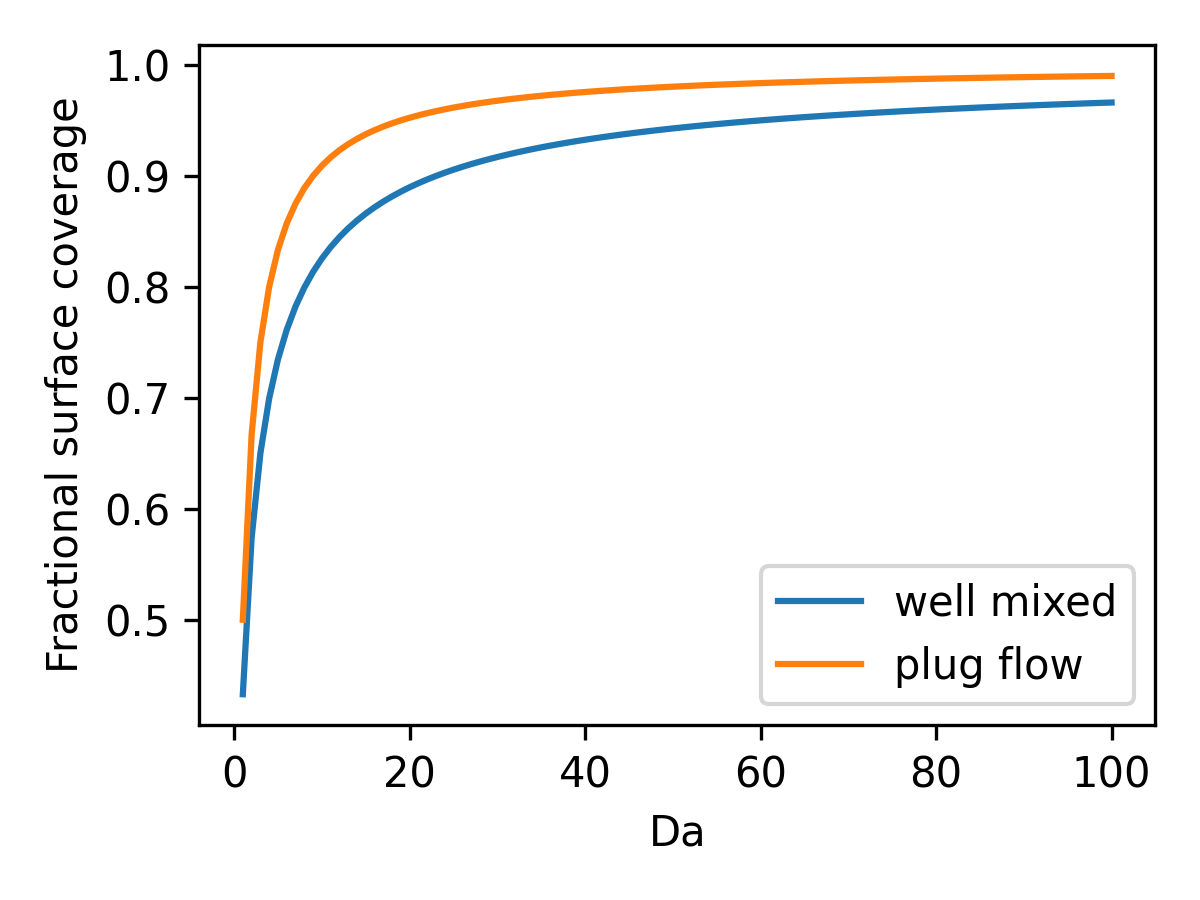}
\caption{\label{fig7} Fractional surface coverage of particles coated by ALD in a continuous process for a normalized particle residence time $\tau_s=1$ as a function of the Damköhler number. As in the case of batch processes, reactors whose precursor transport can be approximated by a plug flow model are more efficient. }
\end{figure}

In Figure \ref{fig7} we compare the fractional surface coverage achieved at $\tau_s=1$ in the well-mixed and plug flow continuous models. As in the case of batch processes, continuous processes where precursor transport can be approximated with a plug flow model consistently lead to higher fractional surface coverages across the whole range of $\mathrm{Da}$ explored. This highlights the importance of reactor design and precursor delivery to ensure a faster transition to a transport limited regime and an optimal precursor utilization.

\subsection{Extension to soft-saturating processes}

The results obtained thus far have focused on an ideal irreversible first order Langmuir surface kinetics. When we extend the models to the soft-saturating case, we observe similar trends, with the plug-flow approximation being consistently faster in transitioning to a transport limited regime and achieving high precursor utilization than the well-mixed approximation for both continuous and batch processes.  

\begin{figure}
\includegraphics[width=8cm]{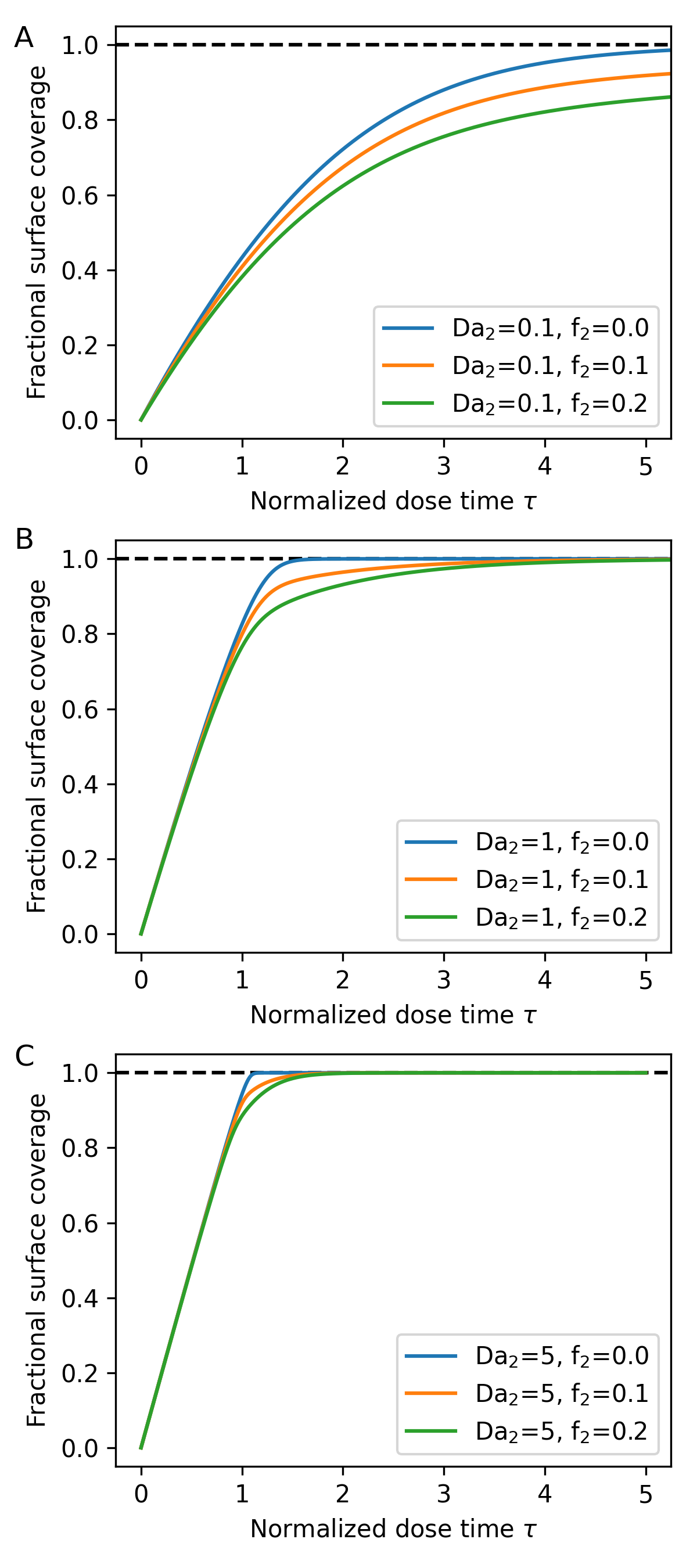}
\caption{\label{fig8} Saturation curves in a well-mixed batch process for a soft-saturating self-limited
surface kinetics. Results are shown for increasing values of the Damköhler number of the main surface reaction pathway: (A) $\mathrm{Da}_1=1$; (B) $\mathrm{Da}_1=10$; (C)$\mathrm{Da}_1=50$. In all cases, $\mathrm{Da}_2=0.1\times \mathrm{Da}_1$. }
\end{figure}

As an example of a soft-saturating process, we have considered a system where the second pathway is ten times less reactive than the main reaction pathway. Based on Eq. \ref{eq:Da}, this means that $\mathrm{Da}_2=0.1\times \mathrm{Da}_1$. In Figure \ref{fig8}, we show the saturation profiles for increasing values of $\mathrm{Da}_1$  for the well-mixed batch process. The presence of a second, slower reaction pathway softens the transition from reaction limited to transport limited regime, with the soft saturating component pushing the transition to higher values of $\mathrm{Da}_1$  due to the presence of the secondary reaction pathway with a lower reactivity. The behavior observed for this model is representative of the plug flow and continuous models.

\section{Discussion}

The four models introduced in this work allow us to evaluate the scalability of different strategies for particle coatings by atomic layer deposition by calculating the transition from reaction-limited to transport limited regimes. In all cases, this transition is dictated by the Damköhler number (Eq. \ref{eq:Da}). In terms of the volumetric flow $\phi$ (in m$^3$s$^{-1}$) into the reactor, the Damköhler number is given by:
\begin{equation}
    \mathrm{Da} = \frac{S}{\phi}\beta_0 \frac{1}{4}v_{th}
\end{equation}
Here, all the parameters except for the sticking probability, $\beta_0$, and the fraction of reactive sites $f$ are available experimentally: the mean thermal velocity, $v_{th}$, depends on the precursor molecular mass and process temperature, the volumetric flow, $\phi$, can be calculated from the process pressure and the standard flow in the mass flow controllers, and $S$ is the total surface area of the particles inside the reactor. If only a fraction $f$ of the sites are reactive,
this effect can be incorporated as an effective surface $S_\mathrm{eff}=fS$.

In the transport limited regime, the saturation dose time or the particle residence time is given simply by the time required to introduce into the reactor a number of precursor molecules equal to the number of available surface sites. This will be limited by the maximum precursor partial pressure that can be delivered at a volumetric flow, $\phi$. It also depends on the number of surface sites per unit area, which can be extracted from the growth per cycle of the ALD process. Consequently, if we assume that $f=1$, the sticking probability is the only parameter that is generally not known in an ALD process.

The results obtained also emphasize the importance of reactor design in ensuring a fast and efficient transition from a reaction-limited to a transport-limited regime: the plug flow model consistently leads to faster processes and higher precursor utilization in both batch and continuous ALD configurations. In the batch process case, fluidized bed reactors are one way of achieving this cross flow. In the continuous case, plug flow configurations show additional properties, such as a self-extinguishing behavior: as shown in Figure \ref{fig6}, the process conditions can be optimized to achieve saturation while minimizing the fraction of unreacted precursor. This can prove beneficial in spatial ALD configurations where isolation between different precursor zones is key.
We have also observed that, in the case of soft-saturating processes where there is a secondary slower
reaction pathway,  the transition from reaction to a transport limited regime shifts to higher Damköhler numbers. However, as shown in Figure \ref{fig8}, running processes in a sub-saturating regime can provide a viable tradeoff between throughput and precursor utilization.

Finally, there is the question of how well these models agree with experiments.
As discussed in Section \ref{sec_batch}, for the case of fluidized reactors we have observed good
qualitative and quantitative agreement between the plug flow model and experimental results of particle coating by TMA/water\cite{Lu_fluidizedbedALD_2022}. The punch-through of the precursor signal as
measured by mass spectrometry, the near 100\% precursor consumption prior to reaching full
saturation, and even the predicted saturation dose times were found to agree well with the model
predictions. However, further experimental work is needed in order to validate the approximations
in other promising reactor configurations. The models explored in this work also don't consider
additional sources of non-idealities that can affect process scale up. One example is particle agglomeration,
where the need for the precursor to diffuse within the agglomerates can lead to longer saturation times.
These effects require more detailed simulation approaches that are beyond the scope of this work\cite{Grillo_multiscalefluidized_2015,Jin_fluidizedALDsim_2017}.

\section{Conclusions}

In this work, we have explored the scalability of different strategies for particle coating by atomic layer deposition by analyzing the transition from reaction-limited to transport-limited regime. We have introduced four simple models encompassing both batch and continuous processes of particle coating with agitation. These models depend primarily on inputs available from either knowledge about the ALD process or the experimental conditions. They can help evaluate the scalability of different ALD processes and their implementation at manufacturing scale, something relevant for cost-sensitive applications such as catalysis, energy storage, and decarbonization. The models are available as part of the Python package aldsim, and made available in the GitHub repository: https://github.com/aldsim/aldsim.

\begin{acknowledgments}
This research is based upon work supported by  Laboratory Directed Research and Development (LDRD) funding from Argonne National Laboratory, provided by the Director, Office of Science, of the U.S. Department of Energy under Contract No. DE-AC02-06CH11357.  This work was supported in part by the Israel-U.S. Collaborative Water-Energy Research Center (CoWERC), supported by the Binational Industrial Research and Development Foundation under Energy Center grant EC-15.
\end{acknowledgments}

\section*{Author Declarations}

\subsection*{Conflict of interest}

The authors have no conflicts to disclose

\section*{Data Availability Statement}

The data that support the findings of
this study are openly available in
the GitHub repository: https://github.com/aldsim/aldparticle.

\section*{References}

\bibliography{particle}

\appendix

\section{Non-dimensional equations}

This document includes the non-dimensional equations for the eight models considered in the manuscript.

\subsection{Batch process, well-mixed}

Particles are homogeneously mixed, and precursor transport is modeled using the
well-mixed approximation (Section \ref{sec:modelA}).

\subsubsection{Single reaction pathway}

\begin{eqnarray}
(1-x) & = & \mathrm{Da} (1-\Theta) x \\
\frac{d \Theta}{d\tau} & = & \mathrm{Da} (1-\Theta) x
\end{eqnarray}
where:
\begin{eqnarray}
x & = & n/n_0 \label{eq_x}\\
\tau & = & t/t_0 \label{eq_tau}\\
\mathrm{Da} & = &  \frac{S}{V} f \beta_0 \frac{1}{4} v_{th} t_\mathrm{res} \label{eq_Da}\\
t_0 & = & t_\mathrm{res}\frac{S}{s_0 n_0 V} \label{eq_t0}
\end{eqnarray}

\subsubsection{Soft-saturating process}

\begin{equation}
\begin{split}
(1-x)  = & \left[(1-f) \mathrm{Da}_a(1-\Theta_a) \right.\\
& \left. + f  \mathrm{Da}_b(1-\Theta_b) \right] x
\end{split}
\end{equation}
and
\begin{eqnarray}
\frac{d \Theta_a}{d\tau} & = & \mathrm{Da}_a (1-\Theta_ b) x \\
\frac{d \Theta_b}{d\tau} & = &\mathrm{Da}_b (1-\Theta_ b) x
\end{eqnarray}
where $x$, $\tau$ and $t_0$ are defined in Eqs. \ref{eq_x}, \ref{eq_tau}, and \ref{eq_t0} and:
\begin{eqnarray}
\mathrm{Da}_a & = & \frac{S}{V}  \beta_a \frac{1}{4} v_{th} t_\mathrm{res} \label{eq_Daa}\\
\mathrm{Da}_b & = & \frac{S}{V}  \beta_b \frac{1}{4} v_{th} t_\mathrm{res} \label{eq_Dab}
\end{eqnarray}

\subsection{Batch process, plug flow}

Particles are homogeneously mixed, and the precursor transport is modeled using a plug flow
model. This corresponds to situations where the precursor is forced to flow through volatilized
particles (Section \ref{sec:modelB}).

\subsubsection{Single reaction pathway}

\begin{eqnarray}
\frac{d x}{d\xi}& = & -  \mathrm{Da} (1-\Theta) x \\
\frac{d \Theta}{d\tau} & = & \mathrm{Da} (1-\Theta) \bar{x} \\
\bar{x} & =&   \int_0^1 x(\xi) d\xi
\end{eqnarray}
where $x$, $\tau$, $\mathrm{Da}$, and $t_0$ are defined in Eqs. \ref{eq_x}, \ref{eq_tau}, \ref{eq_Da}, and \ref{eq_t0} with
\begin{equation}
t_\mathrm{res} = \frac{L}{u} \label{eq_tres}
\end{equation}
and
\begin{equation}
\xi = \frac{z}{L} \label{eq_xi}
\end{equation}

\subsubsection{Soft-saturating process}

\begin{equation}
\begin{split}
\frac{d x}{d\xi} = & - \left[(1-f) \mathrm{Da}_a(1-\Theta_a) \right. \\
  & + \left. f  \mathrm{Da}_b(1-\Theta_b) \right] x
\end{split}
\end{equation}
and
\begin{eqnarray}
\frac{d \Theta_a}{d\tau} & = & \mathrm{Da}_a (1-\Theta_ b) \bar{x} \\
\frac{d \Theta_b}{d\tau} & = &\mathrm{Da}_b (1-\Theta_ b) \bar{x} \\
\bar{x} & =&   \int_0^1 x(\xi) d\xi
\end{eqnarray}
with $\mathrm{Da}_a$ and  $\mathrm{Da}_b$ defined in Eqs. \ref{eq_Daa} and \ref{eq_Dab}.

\subsection{Continuous process, well-mixed}

Particles move along the reactor and homogeneously mixed in the plane perpendicular to the
direction of movement and stratified in the direction of movement. Precursor transport is modeled
using the well-mixed approximation (Section \ref{sec:modelC}).

\subsubsection{Single reaction pathway}

\begin{eqnarray}
(1-x) & = & \mathrm{Da} (1-\Theta) x \\
\frac{d \Theta}{d\xi} & = & \tau_s \mathrm{Da} (1-\Theta) x \\
\bar{\Theta} & = &   \int_0^1 \Theta(\xi) d\xi
\end{eqnarray}

Here:
\begin{equation}
\tau_s = \frac{L}{v t_0} \label{eq_taus}
\end{equation}
and $x$, $\tau$, $\mathrm{Da}$, $t_0$ and $\xi$ are defined in Eqs. \ref{eq_x}, \ref{eq_tau}, \ref{eq_Da}, and \ref{eq_t0}, and \ref{eq_xi}.

\subsubsection{Soft-saturating process}

\begin{equation}
\begin{split}
(1-x) & = \left[(1-f) \mathrm{Da}_a(1-\bar{\Theta}_a)\right. \\
 & \left. + f  \mathrm{Da}_b(1-\bar{\Theta}_b) \right] x
\end{split}
\end{equation}
and

\begin{eqnarray}
\frac{d \Theta_a}{d\xi} & = & \tau_s \mathrm{Da}_a (1-\Theta_ b) x \\
\frac{d \Theta_b}{d\xi} & = &\tau_s \mathrm{Da}_b (1-\Theta_ b) x
\end{eqnarray}
with
\begin{eqnarray}
\bar{\Theta}_a & = &   \int_0^1 \Theta(\xi) d\xi \\
\bar{\Theta}_b & = &   \int_0^1 \Theta(\xi) d\xi
\end{eqnarray}

\subsection{Continuous process, plug flow}

Particles move along the reactor through a continuous feed system. Particles
are homogeneously mixed in the plane perpendicular to the direction of movement and stratified in the direction of movement. Precursor moves downstream along
the direction of the particles and its transport is modeled
using a plug flow approximation (Section \ref{sec:modelD}).

\subsubsection{Single reaction pathway}

\begin{eqnarray}
\frac{d x}{d\xi}& = & -  \mathrm{Da} (1-\Theta) x \\
\frac{d \Theta}{d\xi} & = & \tau_s \mathrm{Da} (1-\Theta) x 
\end{eqnarray}

\subsubsection{Soft-saturating process}

\begin{equation}
\begin{split}
\frac{d x}{d\xi}= & - \left[(1-f) \mathrm{Da}_a(1-\Theta_a) \right. \\
& \left. + f  \mathrm{Da}_b(1-\Theta_b) \right] x
\end{split}
\end{equation}
and

\begin{eqnarray}
\frac{d \Theta_a}{d\xi} & = & \tau_s \mathrm{Da}_a (1-\Theta_ b) x \\
\frac{d \Theta_b}{d\xi} & = & \tau_s \mathrm{Da}_b (1-\Theta_ b) x
\end{eqnarray}

\end{document}